\documentclass[prl,twocolumn,showpacs,amsmath,amssymb,nofootinbib]{revtex4}

\usepackage{epsfig}
\usepackage{graphicx}
\usepackage{amsfonts}
\usepackage{amssymb}
\usepackage{bm}
\usepackage{latexsym}
\usepackage{amsmath}
\usepackage{hyperref}

\newcommand{\beq}{\begin{equation}}
\newcommand{\eeq}{\end{equation}}
\newcommand{\bea}{\begin{eqnarray}}
\newcommand{\eea}{\end{eqnarray}}
\newcommand{\p}{\partial}
\newcommand{\ep}{\epsilon}
\newcommand{\vect}[1]{\bm{\mathrm{{#1}}}}

\begin{document}


\title{Consistency relation for cosmic magnetic fields}

\author{Rajeev Kumar Jain$^{1}$ and Martin S. Sloth$^{2}$}
\affiliation{$^1$D\'{e}partment de Physique Th\'{e}orique and Center for Astroparticle Physics,
\vskip 0.01cm
Universit\'{e} de Gen\`{e}ve, 24 Quai E. Ansermet, CH-1211 Gen\`{e}ve 4, Switzerland
\vskip 0.1cm
$^2$CP$^3$-Origins, Centre for Cosmology and Particle Physics Phenomenology, 
\vskip 0.01cm
University of Southern Denmark, Campusvej 55, 5230 Odense M, Denmark
}


\begin{abstract}
If cosmic magnetic fields are indeed produced during inflation, they are likely to be correlated with the scalar metric perturbations that are 
responsible for the Cosmic Microwave Background anisotropies and Large Scale Structure. Within an archetypical model of inflationary 
magnetogenesis, we show that there exists a new simple consistency relation for the non-Gaussian cross correlation function of the scalar 
metric perturbation with two powers of the magnetic field in the squeezed limit where the momentum of the metric perturbation vanishes. 
We emphasize that such a consistency relation turns out to be extremely useful to test some recent calculations in the literature. Apart from 
primordial non-Gaussianity induced by the curvature perturbations, such a cross correlation might provide a new observational probe of inflation 
and can in principle reveal the primordial nature of cosmic magnetic fields.
\end{abstract}


\maketitle


Cosmological inflation has become a very successful and dominant paradigm to understand the initial conditions for the Cosmic Microwave 
Background (CMB) anisotropies and Large Scale Structure formation~\cite{lrreview}. In particular, the precise observations by the 
Wilkinson Microwave Anisotropy Probe have provided a tremendous support for the inflationary paradigm~\cite{wmap7}. These 
observations are remarkably consistent with the predictions of minimal single field slow-roll inflationary models and in order to test further, it has 
become really important to look for new observables, such as non-Gaussianity that can be used to discriminate between different models of 
inflation and possibly rule out the minimal model. The detection of non-Gaussianity has therefore become one of the primary goals of present 
and future experiments~\cite{planck}. But it is very model dependent, whether primordial non-Gaussianity from inflation will be observable.

Another possibility which has been contemplated is that the inflaton, or some other light field during inflation, for example, the curvaton~\cite{Enqvist:2001zp} is coupled directly to electromagnetism. Such a coupling breaks the conformal invariance of electromagnetism and as a result, large 
scale magnetic fields can be produced during inflation~\cite{Ratra:1991bn}. It has been speculated that this could even provide the seed fields 
required for the galactic dynamo~\cite{Bamba:2003av} and also lead to acoustic signatures in the CMB \cite{Yamazaki:2011eu,Seery:2008ms}.
However, there exists a theoretical problem for such models to succeed, called the strong coupling problem~\cite{Demozzi:2009fu}. It is a remarkably serious problem and as 
discussed in~\cite{Barnaby:2012tk}, it is very difficult to avoid it while retaining gauge invariance unless one gives up on the gauge field being 
the actual electromagnetic field, and instead thinks of it as some novel hidden sector isocurvature gauge field. Recently, it has also been 
speculated that the strong coupling problem can be circumvented, if gauge invariance is broken in the UV in the effective four dimensional 
Lagrangian, but manifest in five dimensions~\cite{Motta:2012rn}. One might also worry that the usual ``$\eta$-problem" of inflation could be 
aggravated by such a coupling of the inflaton to the gauge field~\cite{Barnaby:2011qe}. Independent of whether any specific proposal to 
overcome the strong coupling problem may {\it or} may not work, the model is an archetypical model of magnetogenesis, and can be tested at a strictly phenomenological level by checking the magnetic consistency relation developed here. But as we discuss 
below, our motivations to study such a scenario are even more general.

In a model with a direct coupling between electromagnetism and inflaton, the metric fluctuations will be correlated with the produced large scale 
magnetic fields. Apart from primordial non-Gaussianity, the three-point correlation of the metric perturbation with two powers of the magnetic 
field might provide a new observational probe of inflation that could also shed light on the primordial nature of cosmic magnetic fields and has 
been studied recently in~\cite{Caldwell:2011ra,Motta:2012rn}. 
In this letter, we show that in the squeezed limit where the momentum of the scalar metric perturbation vanishes, there exists a simple 
consistency relation that can be used to obtain the three-point correlation function of the metric perturbation with two powers of the magnetic 
field in a simple way which can be used to check the results in the literature. 

In order to evaluate the cross correlation of the curvature perturbation with the magnetic fields, it is convenient to define the magnetic non-linearity parameter $b_{NL}$, in terms of the correlation function of the curvature perturbation with the magnetic fields as
\bea \label{zBB1}
& & \left<\zeta(\tau_I,\vect{k}_1){\bf B}(\tau_I,\vect{k}_2)\cdot {\bf B}(\tau_I,\vect{k}_3)\right> \nonumber \\
& & = b_{NL}(2\pi)^3\delta^{(3)}(\vect{k}_1+\vect{k}_2+\vect{k}_3)P_\zeta(k_1) P_B(k_2) 
\eea
where $P_\zeta$ and $P_B$ are the power spectra of the comoving curvature perturbation and the magnetic fields, respectively and are defined as
\bea
\left<\zeta(\tau,\vect{k})\zeta(\tau,\vect{k}')\right> =(2\pi)^3\delta^{(3)}(\vect{k}+\vect{k}') P_\zeta(k),\\
\left<{\bf B}(\tau,\vect{k})\cdot {\bf B}(\tau,\vect{k}')\right> =(2\pi)^3\delta^{(3)}(\vect{k}+\vect{k}') P_B(k).
\eea
The conformal time $\tau$ is defined by $a d\tau=dt$ where $a(t)$ is the scale factor of the Friedmann-Lema\^{\i}tre-Robertson-Walker metric $ds^2=-dt^2+a^2(t)d\vect x^2$ and $\tau_I$ denotes the conformal time at the end of inflation.

The time-dependent coupling of the electromagnetic field to the background, can be parametrized by a coupling of the form $\lambda(\phi)F_{\mu\nu}F^{\mu\nu}$, where $F_{\mu\nu} \equiv \p_\mu A_\nu-\p_\nu A_\mu$ is the electromagnetic field strength and the time dependence of the coupling is parametrized by its dependence on a slowly rolling background scalar field $\phi$, which we think of as being the inflaton for simplicity.
When $b_{NL}$ is momentum independent, it corresponds to a ``local" non-linearity which can be obtained from the relation
\beq
{\bf B} = {\bf B}^{(G)} +\frac{1}{2}b_{NL}^{local} \zeta^{(G)}  {\bf B}^{(G)}
\eeq
where ${\bf B}^{(G)}$ and $\zeta^{(G)}$ are the Gaussian fields.
One can estimate the size of $b_{NL}$ by noting that the interaction Lagrangian between the scalar field and the electromagnetic field is $\mathcal{L}_{BB} \propto \lambda(\phi)F^2$. By Taylor expanding the coupling in the inflaton fluctuations, $\lambda(\phi) = \lambda(\phi_c)+\p_\phi \lambda(\phi_c)\delta\phi$, one obtains that the linear coupling between the inflaton fluctuation and the electromagnetic field is $\mathcal{L}_{\zeta BB} \propto \p_\phi \lambda(\phi) \delta\phi F^2$. It is useful to express the scalar perturbations in terms of the comoving curvature perturbation $\zeta$ which can be considered as the scalar perturbation of the metric
\beq
ds^2 =-dt^2+a^2(t)\,e^{2 \zeta(t,\vect x)}d\vect x^2
\eeq 
on large scales where the time derivate of $\zeta$ vanishes.
The comoving curvature perturbation is related to the inflation fluctuation by $\delta\phi =\sqrt{2\ep}\,\zeta$ where the slow-roll parameter $\ep$ is given by $\sqrt{2\ep} = -\dot\phi/H$. With these definitions,  we have 
\beq\label{deriv}
\p_\phi\lambda  \delta\phi = \frac{d \lambda}{d t}\frac{d t}{d \phi} \delta\phi =-\dot \lambda \zeta/H
\eeq
which leads to $\mathcal{L}_{\zeta BB} \propto -\dot\lambda (\zeta/H)F^2$. In analogy with the analysis of \cite{Seery:2008ax}, we can compare it with the quadratic term $\mathcal{L}_{BB}$. The ratio is  $\mathcal{L}_{\zeta BB} /\mathcal{L}_{BB}\propto \dot\lambda/(H\lambda)\cdot P_{\zeta}^{1/2}$, and we would therefore expect
\beq
b_{NL} \sim  \mathcal{O}(\dot\lambda/(H\lambda))~.
\eeq

We will show later that in the squeezed limit, where the momentum of the curvature perturbation vanishes i.e. $k_1 \ll k_2,k_3=k$, the magnetic consistency relation in fact gives
\beq\label{bnllocal}
b_{NL}^{local} =  -\frac{1}{H}\frac{\dot\lambda}{\lambda}
\eeq
which agrees with the naive expectations. 

Despite a basic difference (which will be evident later), due to the conformal nature of electromagnetism when the coupling $\lambda$ is trivial, the consistency relation shares some familiarity with other consistency relations, which have previously proven to be extremely powerful in the literature. In~\cite{Maldacena:2002vr,Creminelli:2004yq}, a consistency relation of the bispectrum of the curvature perturbation was obtained, and in~\cite{Seery:2008ax}, a new consistency check of the tree-level exchange diagrams, where a scalar or a graviton is propagating between two pairs of external legs, was derived. Some of these consistency relations have been extended to higher orders using arguments related to conformal symmetries in~\cite{Creminelli:2012ed,Kehagias:2012pd}. Finally, the semiclassical relation for loop diagrams in~\cite{Giddings:2010nc,Giddings:2011zd} made it possible to clarify the nature of IR loop effects in de Sitter. Some of these approaches are reviewed in the appendix of~\cite{Giddings:2010nc}.

In the Coulomb gauge with $A_0=0$ and $\p_iA^i=0$, the quadratic action for the electromagnetic vector field $A_i$ becomes
\bea\label{A2}
S_{em} &=& -\frac{1}{4}\int d^4 x \sqrt{-g}\,\lambda(\phi) F_{\mu\nu} F^{\mu\nu} \nonumber \\
&=&\frac{1}{2} \int d^3 x\, d\tau \lambda(\phi) \left({A_i'}^2 -\frac{1}{2}(\p_iA_j-\p_jA_i)^2\right).
\eea
Since the magnetic field is a divergence free vector field, the two-point correlation function of the magnetic field can be written as
\beq\label{BB}
\left< B_i(\tau,\vect{k}){B^i}(\tau,\vect{k}')\right> = \frac{k^2}{a^4}\left(\delta_{ij}-\frac{ k_i k_j}{k^2}\right) \left< A_i(\tau,\vect{k}) A_j(\tau,\vect{k}') \right>
\eeq
and the magnetic field power spectrum becomes 
\beq
 P_B(k)=\frac{k^2}{a^4}\left< {\bf A}(\tau,\vect{k})\cdot {\bf A}(\tau,-\vect{k}) \right>.
\eeq
By parametrizing the time-dependence of the coupling function as 
\beq\label{lambda}
\lambda(\phi(\tau)) = \lambda_I (\tau/\tau_I)^{-2n}~, 
\eeq
one finds that the spectral index, $n_B$, of the magnetic field energy density, $d\rho_B/d\log k$, is given by $n_B=4-2n$ for $n\geq 0$ \cite{Bamba:2003av}.

We are therefore interested in computing a basic correlation function such as $\left<\zeta(\tau_I,\vect k_1) A_i(\tau_I,\vect k_2) A_j(\tau_I,\vect k_3)\right>$ in the squeezed limit $k_1 \ll k_2,k_3$. Since $\zeta$ is frozen outside the horizon, the effect of $\zeta_{k_1}$ in the squeezed limit is to locally rescale the background when computing the correlation functions on shorter scales given by $k_2, k_3$ and one can write~\cite{Maldacena:2002vr,Creminelli:2004yq}
\bea\label{rel1}
& & \lim_{k_1 \to 0} \left< \zeta(\tau_I,\vect k_1) A_i(\tau_I,\vect k_2) A_j(\tau_I,\vect k_3)\right>\nonumber \\ 
& & = \left< \zeta(\tau_I,\vect k_1)\left< A_i(\tau_I,\vect k_2) A_j(\tau_I,\vect k_3)\right>_B\right>,
\eea
where $\left<A_i(\tau_I,\vect k_2) A_j(\tau_I,\vect k_3)\right>_B$ is the correlation function of the short wavelength modes due to the variation in the background produced by the long wavelength mode of $\zeta$.

In the absence of the coupling function $\lambda$, the conformal invariance of the gauge field implies that it only feels the background expansion through $\lambda$ which subsequently depends on the scale factor. Since $A_i(\tau,\vect k)$ doesn't feel the background for a trivial $\lambda$ and the only effect of $\zeta(\tau,\vect k)$ in the squeezed limit is to locally rescale the background as $a\to a_B=e^{\zeta_B}a$, we expect that the correlation function $\left<\zeta(\tau_I,\vect k_1) A_i(\tau_I,\vect k_2) A_j(\tau_I,\vect k_3)\right>$ will vanish in this limit to the leading order for a trivial $\lambda$.

To compute the correlation function for a non-trivial $\lambda$, we need to write $A_i(\tau,\vect k)$ in terms of the Gaussian one with a trivial $\lambda$. If we expand $\lambda_B = \lambda(a_B)$ around a homogenous background value $\lambda_0=\lambda(a)$, we get
\beq\label{dN}
\lambda_B = \lambda_0 + \frac{d \lambda_0}{d\ln a}\delta\ln a +\dots = \lambda_0+\frac{d \lambda_0}{d\ln a} \zeta_B+\dots ~.
\eeq
Notice the resemblance with a $\delta N$ expansion. Using $N=\ln a$, we could also have written the above as $\lambda=\lambda_0 +\lambda'_N\delta N +\dots$, where $\lambda'_N \equiv \p \lambda/\p N$.

Defining the linear Gaussian part of $A_i$ to be $A_i^{(G)}$ and in the Coulomb gauge with a pump field $S^2 =\lambda_0$, we can define a linear Gaussian canonical vector potential $v_i = S(\tau) A^{(G)}_i$, such that the quadratic action in~(\ref{A2}) takes the form similar to a canonical scalar field with an effective time dependent mass term $S''/S$ and can be written as
\beq\label{actvc} 
S_{v} =   \frac{1}{2}\int d\tau d^3 x  \left[v_i'^2-(\p_j v_i)^2+\frac{S''}{S}v_i^2\right].
\eeq
In the squeezed limit, all the effect of $\zeta_B$ on the gauge field is captured by~(\ref{dN}) and therefore, we have
\bea
& & \left< A_i(\tau,\vect x_2) A_j(\tau,\vect x_3)\right>_B = \left<  \frac{1}{\lambda_B} v_i(\tau, \vect x_2)v_j(\tau,\vect x_3)\right> \nonumber\\
& &\qquad\qquad\qquad \simeq \frac{1}{\lambda_0} \left<v_i(\tau,\vect x_2) v_j(\tau,\vect x_3)\right> \nonumber  \\
& &\qquad\qquad\qquad -\frac{1}{\lambda_0^2}\frac{d \lambda}{d\ln a}\zeta_B \left<  v_i(\tau,\vect x_2) v_j(\tau,\vect x_3)\right>
\eea
which leads to
\bea
& &\left< A_i(\tau,\vect x_2) A_j(\tau,\vect x_3)\right>_B \simeq \left< A_i(\tau,\vect x_2) A_j(\tau,\vect x_3)\right>_0 \nonumber \\
& &\qquad\qquad~ - \frac{1}{\lambda_0^2}\frac{d \lambda}{d\ln a} \zeta_B\left< A_i(\tau,\vect x_2)A_j(\tau,\vect x_3)\right>_0
\eea
and a Fourier transformation gives
\bea
& &\!\!\!\!\!\!\!\!\!\! \left<A_i(\tau, \vect k_2) A_j(\tau, \vect k_3)\right>_B \simeq \left<A_i(\tau, \vect k_2) A_j(\tau, \vect k_3)\right>_0 \nonumber \\
& &\!\!\!\!\!\!\!\! -\frac{1}{\lambda_0}\frac{d\lambda}{d\ln a}\int \frac{d^3 k_B}{(2\pi)^{3}}\zeta(\tau, \vect k_B)\left<A_i(\tau, \vect k_2) A_j(\tau, \vect k_3)\right>_0
\eea
Using $d\lambda/d\ln a = \dot\lambda/H$, we then find from~(\ref{rel1}) the squeezed limit consistency relation for the gauge field
\bea
& &\lim_{k_1 \to 0} \left< \zeta(\tau_I,\vect k_1) A_i(\tau_I,\vect k_2) A_j(\tau_I,\vect k_3)\right> \nonumber \\
&&   \simeq -\frac{1}{H}\frac{\dot\lambda}{\lambda}\left<\zeta(\tau_I,\vect k_1)\zeta(\tau_I,-\vect k_1)\right>_0 \left<A_i(\tau_I, \vect k_2) A_j(\tau_I, \vect k_3)\right>_0 \nonumber \\
\eea
 Finally, by using the relation (\ref{BB}), we find the consistency relation for the magnetic field to be
\bea \label{zBB2}
& &\!\!\!\!\!\!\!\!\!\! \left<\zeta(\tau_I,\vect{k}_1){\bf B}(\tau_I,\vect{k}_2)\cdot {\bf B}(\tau_I,\vect{k}_3)\right> \nonumber \\
& &\!\!\!\!\!\!\!\!\! = -\frac{1}{H}\frac{\dot\lambda}{\lambda}(2\pi)^3\delta^{(3)}(\vect{k}_1+\vect{k}_2+\vect{k}_3)P_\zeta(k_1) P_B(k_2)
\eea
in agreement with~(\ref{bnllocal}). With the parametrization in~(\ref{lambda}), we obtain the consistency relation $b_{NL}= n_B-4$. For the most interesting case of a scale invariant spectrum, $n_B=0$, the magnetic non-linearity parameter, $b_{NL}$, is non-vanishing.

In the squeezed limit, the consistency relation is quite general as an explicit form of the coupling function has not been used. But as we argue below, the approximation used to obtain this consistency relation might only be trusted for $n\geq1$ in~(\ref{lambda}). To see this, note that for a canonical massless scalar field in de Sitter space, the pump field in~(\ref{actvc}) can be identified with the scale factor, and one would have $S''/S =a''/a$ with $a\propto 1/\tau$. In de Sitter, there exists an apparent future event horizon and therefore, the long and the short wavelength modes decouple on super horizon scales. As a result, all the effects of the long wavelength modes can be captured by their rescaling of the background of the short wavelength modes. In the more general case above, if we interpret $S$ as an effective scale factor for the background experienced by $v_i$, it is evident that this effective background has an apparent future event horizon {\it only} for $n\geq 1$ and we can trust the decoupling of long wavelength modes from short wavelength modes. On the other hand, for $n\leq-2$, we can use the symmetry of $S''/S=n(n+1)/\tau^2$ under $n\to -(n+1)$ to obtain a consistency relation for those values, once we have a consistency relation valid for $n\geq1$.

It is interesting to compare our result in~(\ref{zBB2}), with the calculations of~\cite{Motta:2012rn} and~\cite{Caldwell:2011ra}. In~\cite{Caldwell:2011ra}, the comoving curvature perturbation was replaced with the perturbation of a test scalar field. On the other hand, if we use the relation $\zeta = \delta\phi/\sqrt{2\ep}$, the dominant interaction of the metric perturbations with the electromagnetic field can be obtained by Taylor expanding the coupling in the inflaton fluctuations as $\lambda(\phi) = \lambda(\phi_c)+\p_\phi \lambda(\phi_c)\delta\phi$. In this way, one obtains the linear coupling between the inflation fluctuation and the electromagnetic field $\mathcal{L}_{\delta\phi BB} =-(1/4) \p_\phi \lambda(\phi) \delta\phi F^2$. This is same as the coupling considered in~\cite{Caldwell:2011ra} and we are able to reproduce their results in the squeezed limit when using $\dot \lambda\zeta = -\p_\phi\lambda H \delta\phi$ in order to write the consistency relation as
\bea \label{zBB3} 
& &\!\!\!\!\left<\delta\phi(\tau_I,\vect{k}_1){\bf B}(\tau_I,\vect{k}_2)\cdot {\bf B}(\tau_I,\vect{k}_3)\right> \nonumber \\
& &\!\!= \frac{\p_\phi\lambda}{\lambda}(2\pi)^3\delta^{(3)}(\vect{k}_1+\vect{k}_2+\vect{k}_3)P_{\delta\phi}(k_1) P_B(k_2) 
\eea
and inserting the specific form of the coupling $\lambda(\phi) = \exp(2\phi/M)$ used there. The power spectrum of the scalar field fluctuations is defined as $\left<\delta\phi(\tau,\vect{k})\delta\phi(\tau,\vect{k}')\right> =(2\pi)^3\delta^{(3)}(\vect{k}+\vect{k}') P_{\delta\phi}(k)$. 
By means of numerical checks, it was observed that in the squeezed limit, the final results are independent of $n$~\cite{Caldwell:2011ra}. We can now understand straightforwardly from~(\ref{zBB3}) why this has to be the case.

In~\cite{Motta:2012rn}, the correlation function in~(\ref{zBB2}) was calculated for the actual metric perturbation. The form of the action that they used is slightly more complicated than $\mathcal{L}_{\delta\phi B B} =-(1/4) \p_\phi \lambda(\phi) \delta\phi F^2$ to the leading order in slow-roll. However, it can be shown that the extra terms in their interaction Hamiltonian are proportional to a total derivative~\cite{rm} and in fact, using a full gauge fixing in the Arnowitt-Deser-Misner formalism~ \cite{Arnowitt:1960es}, it was shown in~\cite{Seery:2008ms} that at linear order, the inflaton fluctuation  $\delta\phi$ is related to the comoving curvature perturbation by $\zeta=\delta\phi/\sqrt{2\ep}$ and the relevant part of the action to leading order in slow-roll is indeed $\mathcal{L}_{\delta\phi B B} =-(1/4) \p_\phi \lambda(\phi) \delta\phi F^2$. Therefore, the squeezed limit result of~\cite{Motta:2012rn} should agree with that of~\cite{Caldwell:2011ra}, when using $\zeta = \delta\phi/\sqrt{2\ep}$. This is not what~\cite{Motta:2012rn} finds which indicates that there is numerical factor wrong in the results of~\cite{Motta:2012rn}. We have therefore carefully checked the results of~\cite{Motta:2012rn} and a detailed calculation shows that the correct calculation agrees with the results of~\cite{Caldwell:2011ra} and with the consistency relation in the squeezed limit~\cite{rm}.

We have thus shown in~(\ref{zBB2}) that in a generic model where magnetic fields are generated during inflation, there exists a new type of consistency relation for the three point cross correlation function of the comoving curvature perturbation with two powers of the magnetic field in the squeezed limit where the momentum of the curvature perturbation vanishes. Moreover, this consistency relation can also be written for the correlation of inflaton field fluctuations with the magnetic field as in~(\ref{zBB3}). Similarly, when the electromagnetic field is coupled to a curvaton-like scalar field (instead of the inflaton) ignoring gravity in a fixed de Sitter background, the correlation of the field fluctuation with the magnetic field in the squeezed limit is again given by~(\ref{zBB3}).

It is beyond the scope of the present paper to discuss in detail the observational prospects of our results. Nevertheless, one might note that the three-point cross correlation function has been argued to be observable through the combined measurement of Faraday rotation and large scale structure \cite{Stasyszyn:2010kp,Caldwell:2011ra}.

Although the underlying model suffers from the strong coupling problem~\cite{Demozzi:2009fu,Barnaby:2012tk}, it is an archetypical model of primordial magnetogenesis which has been widely studied in the literature. It will be interesting to see if the strong coupling problem can be avoided (perhaps along the lines suggested in~\cite{Motta:2012rn}), and also, if the results obtained here can be applied to other models. For instance, the gauge field might instead play the role of a vector curvaton leading to some amount of statistical anisotropy and anisotropic non-Gaussianity though it requires considerable fine tuning of the gauge coupling~\cite{Dimopoulos:2011ym}.


{{\it Acknowledgments.}~We would like to thank Ruth Durrer, Nemanja Kaloper and Antonio Riotto for useful discussions. RKJ acknowledges financial support from the Swiss National Science Foundation and thanks the CP$^3$-Origins, Centre for Cosmology and Particle Physics Phenomenology at the University of Southern Denmark for hospitality where part of this work was carried out. MSS is supported by a Jr. Group Leader Fellowship from the Lundbeck Foundation.}



\begin{thebibliography}{99}

\bibitem{lrreview} For a review, see
D.~H.~Lyth and A.~Riotto,
Phys.\ Rept.\  {\bf 314}, 1 (1999). 


\bibitem{wmap7} 
E.~Komatsu {\it et al.}  [WMAP Collaboration],
Astrophys.\ J.\ Suppl.\  {\bf 192}, 18 (2011)


\bibitem{planck} 
See {\it http://planck.esa.int/}.


\bibitem{Enqvist:2001zp}
K.~Enqvist and M.~S.~Sloth,
Nucl.\ Phys.\  B {\bf 626}, 395 (2002);
D.~H.~Lyth and D.~Wands,
Phys.\ Lett.\  B {\bf 524}, 5 (2002);
T.~Moroi and T.~Takahashi,
Phys.\ Lett.\  B {\bf 522}, 215 (2001)
[Erratum-ibid.\  B {\bf 539}, 303 (2002)].


\bibitem{Ratra:1991bn} 
B.~Ratra,
Astrophys.\ J.\  {\bf 391}, L1 (1992);
M.~S.~Turner and L.~M.~Widrow,
Phys.\ Rev.\ D {\bf 37}, 2743 (1988);
L.~M.~Widrow,
Rev.\ Mod.\ Phys.\  {\bf 74}, 775 (2002).


\bibitem{Bamba:2003av}
K.~Bamba and J.~Yokoyama,
Phys.\ Rev.\ D {\bf 69}, 043507 (2004);
K.~Bamba and M.~Sasaki,
JCAP {\bf 0702}, 030 (2007);
J.~Martin and J.~Yokoyama,
JCAP {\bf 0801}, 025 (2008);
K.~Subramanian,
Astron.\ Nachr.\  {\bf 331}, 110 (2010);
R.~Durrer, L.~Hollenstein and R.~K.~Jain,
JCAP {\bf 1103}, 037 (2011);
C.~T.~Byrnes, L.~Hollenstein, R.~K.~Jain and F.~R.~Urban,
JCAP {\bf 1203}, 009 (2012);
R.~K.~Jain, R.~Durrer and L.~Hollenstein,
arXiv:1204.2409 [astro-ph.CO];
T.~Kahniashvili, A.~Brandenburg, L.~Campanelli, B.~Ratra and A.~G.~Tevzadze,
arXiv:1206.2428 [astro-ph.CO].
 
 
\bibitem{Yamazaki:2011eu} For a review, see (and references therein)
  D.~G.~Yamazaki, K.~Ichiki, T.~Kajino and G.~J.~Mathew,
  Adv.\ Astron.\  {\bf 2010}, 586590 (2010);
  D.~G.~Yamazaki, T.~Kajino, G.~J.~Mathew and K.~Ichiki,
  arXiv:1204.3669 [astro-ph.CO]. 
  
 
\bibitem{Seery:2008ms} 
D.~Seery,
JCAP {\bf 0908}, 018 (2009).
  
  
\bibitem{Demozzi:2009fu}
V.~Demozzi, V.~Mukhanov and H.~Rubinstein,
JCAP {\bf 0908}, 025 (2009).

 
\bibitem{Barnaby:2012tk} 
N.~Barnaby, R.~Namba and M.~Peloso,
Phys.\ Rev.\ D {\bf 85}, 123523 (2012).
  
  
\bibitem{Motta:2012rn} 
L.~Motta and R.~R.~Caldwell,
Phys.\ Rev.\ D {\bf 85}, 103532 (2012).


\bibitem{Barnaby:2011qe} 
N.~Barnaby, E.~Pajer and M.~Peloso,
Phys.\ Rev.\ D {\bf 85}, 023525 (2012).


\bibitem{Caldwell:2011ra} 
R.~R.~Caldwell, L.~Motta and M.~Kamionkowski,
Phys.\ Rev.\ D {\bf 84}, 123525 (2011).
  
  
\bibitem{Seery:2008ax} 
D.~Seery, M.~S.~Sloth and F.~Vernizzi,
JCAP {\bf 0903}, 018 (2009).
    

\bibitem{Maldacena:2002vr} 
J.~M.~Maldacena,
JHEP {\bf 0305}, 013 (2003).
  
  
\bibitem{Creminelli:2004yq} 
P.~Creminelli and M.~Zaldarriaga,
JCAP {\bf 0410}, 006 (2004).

 
\bibitem{Creminelli:2012ed} 
P.~Creminelli, J.~Norena and M.~Simonovic,
JCAP {\bf 1207}, 052 (2012). 
  

\bibitem{Kehagias:2012pd} 
A.~Kehagias and A.~Riotto,
Nucl.\ Phys.\ B {\bf 864}, 492 (2012).


\bibitem{Giddings:2010nc} 
S.~B.~Giddings and M.~S.~Sloth,
JCAP {\bf 1101}, 023 (2011).
  

\bibitem{Giddings:2011zd}
S.~B.~Giddings and M.~S.~Sloth,
Phys.\ Rev.\ D {\bf 84}, 063528 (2011).
  
  
\bibitem{rm} 
R.~K.~Jain and M.~S.~Sloth,
to appear.


\bibitem{Arnowitt:1960es} 
R.~L.~Arnowitt, S.~Deser and C.~W.~Misner,
Phys.\ Rev.\  {\bf 117}, 1595 (1960).
  
\bibitem{Stasyszyn:2010kp} 
  F.~Stasyszyn, S.~E.~Nuza, K.~Dolag, R.~Beck and J.~Donnert,
  arXiv:1003.5085 [astro-ph.CO].

\bibitem{Dimopoulos:2011ym} 
K.~Dimopoulos, G.~Lazarides and J.~M.~Wagstaff,
JCAP {\bf 1202}, 018 (2012).
  
  
\end{thebibliography}
\end{document}